\documentclass{aa}
\usepackage{graphicx}
\usepackage{txfonts}

\begin{document}
   \title{Collapse and expansion in the bright-rimmed cloud SFO 11NE}

   \subtitle{}

   \author{Mark A.~Thompson \and Glenn J.~White}
   
   \offprints{M.A.~Thompson }

   \institute{Centre for Astrophysics \& Planetary Science,
              School of Physical Sciences,
              University of Kent,
              Canterbury,
              Kent CT2 7NR,
              UK\\
              \email{m.a.thompson@kent.ac.uk; g.j.white@kent.ac.uk}
	       }

   \authorrunning{Thompson \& White}
   \titlerunning{Collapse and expansion in SFO 11NE}
   
   \date{}

   \abstract{ We report the results of a search for the double-peaked blue-skewed infall
   signature in the bright-rimmed cloud core SFO 11NE SMM1. Observations of the
   optically thick HCO$^{+}$ and optically thin H$^{13}$CO$^{+}$ J=3--2 lines reveal
   that there is indeed a characteristic double-peaked line profile, but skewed to the
   red rather than the blue. Modelling of the dust continuum emission and line profiles
   show that the motions within SFO 11NE SMM1 are consistent with a collapsing central
   core surrounded by an expanding outer envelope. We show that the collapse is
   occurring at a similar rate to that expected onto a single solar-mass protostar and
   is unlikely to represent the  large-scale collapse of gas onto the infrared cluster
   seen at the heart of SFO 11NE SMM1. The outer envelope is expanding at a much greater
   rate than that expected for a photoevaporated flow from the cloud surface. The
   modelled expansion is consistent with the bulk cloud re-expansion phase predicted 
   by radiative-driven implosion models of cometary clouds.
   }

   \maketitle
%

\section{Introduction}

Spectroscopic evidence for collapse motions in low-mass star-forming regions has
now become relatively widespread (e.g. Myers et al.~\cite{myers00}; Evans
\cite{evans99}). The classic spectroscopic signature of infall or collapse is an
asymmetric  blue-skewed double-peaked optically thick line profile with
self-absorption at the systemic velocity (Myers et al.~\cite{myers00}). The
interpretation of this profile can be fraught with difficulty: unrelated clouds
located along the line-of-sight can mimic the infall signature (Myers et
al.~\cite{myers00}), molecular outflows add extra complexity to the gas
kinematics (Hogerheijde et al.~\cite{hvdbvl98}), chemical effects such as
depletion  may mask the infall signature completely (Rawlings \& Yates
\cite{ry01}) and  rotation of the molecular core can completely reverse the
blue-skewed profile to  the red (Zhou \cite{zhou95}).  In order to conclusively
demonstrate the presence of infall it is usually necessary to map the emission of both
optically thick and thin lines across the molecular cloud core and compare their
profiles to collapse models (e.g.~Gregersen et al.~\cite{gemm00}; Lee, Myers \&
Tafalla \cite{lmt01}).

Despite these difficulties, collapse motions have been identified in several
protostellar and preprotostellar cores. The objects that have been mainly investigated
so far are Class 0 \& I protostars (Gregersen et al.~\cite{gezc97}; Mardones et
al.~\cite{mmtwb97}) and preprotostellar cores  (Lee, Myers \& Tafalla~\cite{lmt99},
\cite{lmt01}). Most of these objects are located in nearby relatively isolated
low-mass star-forming regions such as the Perseus, Serpens, Taurus and Ophiuchus 
molecular clouds (Mardones et al.~\cite{mmtwb97}). Few star-forming regions associated
with the clustered star-forming mode (Lada, Strom \& Myers \cite{lsm99}) have been
searched for the presence of infall. Two notable exceptions include recent searches 
for collapse motions of gas associated with young stellar clusters themselves
(Williams \& Myers \cite{wm99}; Williams \& Garland \cite{wg02}) and for infall
within bright-rimmed clouds associated with large HII regions (de Vries, Narayanan \&
Snell \cite{dvns02}).

Both types of search report limited success with collapse motions identified in two
young clusters (Cepheus A and NGC 2264) and one bright-rimmed cloud (SFO 18). It is
the latter type of object that we will focus on in this paper. Bright-rimmed clouds
are potential regions of triggered or induced star formation (Sugitani, Fukui \& 
Ogura~\cite{sfo}; Sugitani \& Ogura \cite{so94}), whereby the collapse of the
molecular gas comprising the cloud and the ensuing star formation process is
initiated by the action of an external trigger. In bright-rimmed clouds the
external  trigger is most likely the photoionisation-induced shocks that are driven
into the clouds by the ionisation of their outer layers by nearby OB stars
(e.g.~Elmegreen \cite{e91}). These shocks compress the molecular gas of the clouds,
triggering the collapse of the cloud and forming a dense, possibly quasistatic core at
the cloud centre (Bertoldi \cite{bertoldi}; Lefloch \& Lazareff \cite{ll94},
\cite{ll95}).

However, although de Vries, Narayanan \& Snell (\cite{dvns02}) report the detection of
an infall signature towards the bright-rimmed cloud \object{SFO 18}, it is the
exception rather than the rule in their sample of bright-rimmed clouds. The overall
blue excess of the bright-rimmed cloud sample is less than that of the Class 0 \& I
protostellar cores observed by Mardones et al.~(\cite{mmtwb97}) and Gregersen et
al.~(\cite{gemm00}). This phenomenon is puzzling given the strong velocity gradients
predicted by the Lefloch \& Lazareff (\cite{ll94}) radiative-driven implosion model.
However the  small sample size of de Vries, Narayanan \& Snell (\cite{dvns02}) means
that their result may not be statistically significant. It is also possible that a
flattened temperature gradient across the bright-rimmed clouds, caused by their 
external heating, could mask the classic asymmetric infall signature (de Vries,
Narayanan \& Snell \cite{dvns02}). A wider infall survey of a larger number of
bright-rimmed clouds, coupled with more realistic radiative transfer modelling of
their temperature and velocity gradients is required to address the issue of infall in
these clouds. Such a study would also permit the detailed investigation of the
kinematics of these clouds, which is important in the context of refining and
expanding the existing radiative driven implosion models to take into account star
formation processes  (e.g.~Bertoldi \cite{bertoldi}; Lefloch \& Lazareff \cite{ll94},
\cite{ll95}).

We are currently undertaking such a survey of a statistically significant
sample of bright-rimmed clouds. As a fore-runner to the survey we present the
results of a search for infall toward the bright-rimmed cloud \object{SFO
11NE}. \object{SFO 11NE} is located at the edge of the HII region IC 1848,
approximately 4\arcmin\ NE of the bright-rimmed cloud SFO 11  (Thompson et
al.~\cite{thompson03a}; Sugitani, Ogura \& Pickles \cite{osp02}). SCUBA sub-mm 
continuum and JCMT CO observations reveal a dense core of molecular gas and
dust located at the head of the cometary cloud (\object SFO 11NE SMM1), whilst
2MASS images show  that the dense dust core harbours a cluster of candidate
young stellar objects and protostars (Thompson et al.~\cite{thompson03a}). The
pressure balance and morphology of the cloud cloud suggest that the cloud is
likely to be in the early collapse phase described by Lefloch \& Lazareff
(\cite{ll94}) and is thus a good candidate in which to search for infall.


\section{Observational procedure and results}
\label{sect:obsvns}

We obtained five spectra of the HCO$^{+}$ J=3--2 line at 267.5576 GHz using the
facility A-band heterodyne receiver at the James Clerk Maxwell Telescope
(JCMT\footnote{The JCMT is operated by the Joint Astronomy Centre on behalf of PPARC
for the United Kingdom, the Netherlands Organisation of Scientific Research, and the
National Research Council of Canada.}) during June 2002. The spectra were arranged in
a 5-point cross-shaped grid centred upon the sub-mm continuum peak of \object{SFO 11NE
SMM1} ($\alpha$(2000) = 02$^{\rm h}$ 51$^{\rm m}$ 53\fs6, $\delta$(2000) =
+60\degr07\arcmin00\arcsec). The FWHM beamwidth of the JCMT is $\sim$22\arcsec\ at
this frequency and the cross-shaped grid was spaced by 15\arcsec. A single spectrum of
the H$^{13}$CO$^{+}$ J=3--2 line at 260.2555 GHz at the sub-mm continuum peak position
was also obtained. The HCO$^{+}$ line is usually optically thick towards typical
star-forming regions (e.g.~Gregersen et al.~\cite{gemm00}), whereas the less common
isotopomeric line is usually optically thin. 


All spectra were obtained using frequency-switched mode with a local oscillator
switching offset of 8.2 MHz. 
backend spectrometer was set to its minimum bandwidth of 125 MHz to obtain the maximum
possible velocity resolution of $\sim$ 0.1 km\,s per 94 kHz wide channel. The
integration times for the   HCO$^{+}$ spectra and H$^{13}$CO$^{+}$  spectrum were 20
minutes and 40 minutes respectively, which resulted in an r.m.s.~noise  per 0.1
km\,s$^{-1}$ wide channel of  90 and 68 mK.

\begin{figure} 
\centering
\includegraphics*[scale=0.135,trim=200 250 150 250]{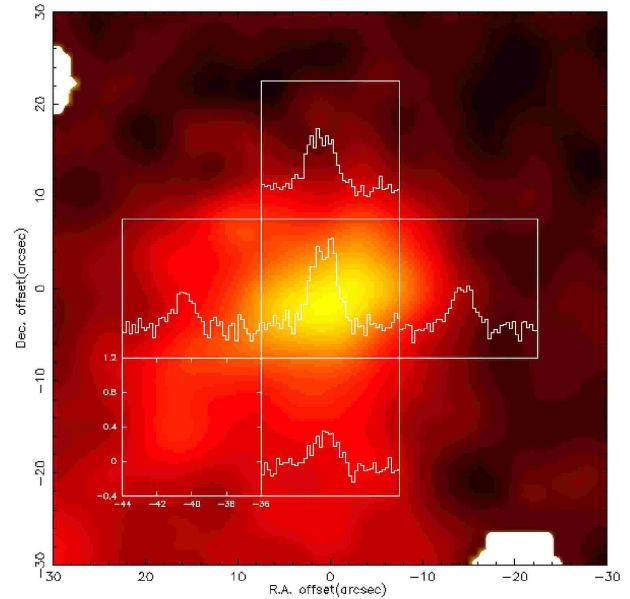}
\caption{Grid-map of the HCO$^{+}$ J=3--2 spectra overlaid on an 850 $\mu$m SCUBA image
of SFO 11NE SMM1 (Thompson et al.~\cite{thompson03a}). The map centre is at 
$\alpha$(2000) = 02$^{\rm h}$ 51$^{\rm m}$ 53\fs6, $\delta$(2000) = +60\degr07\arcmin00\arcsec.}
\label{fig:gridspec}
\end{figure}

\begin{figure}
\centering
\includegraphics*[scale=0.6,angle=-90]{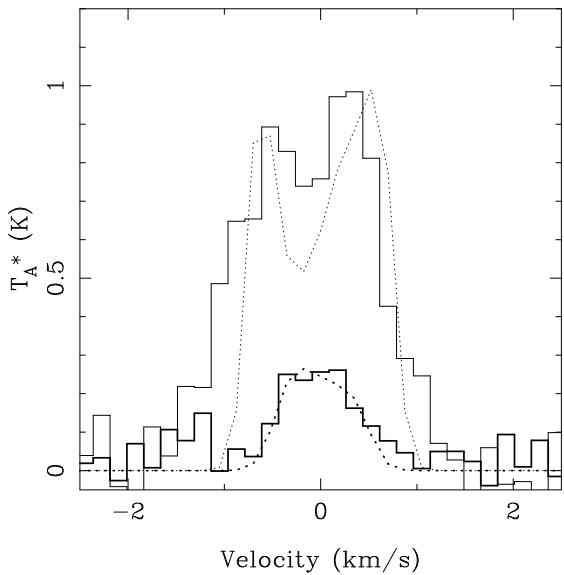}
\caption{Composite spectrum of the HCO$^{+}$  (\emph{thin}) and the
H$^{13}$CO$^{+}$ (\emph{thick}) J=3--2 lines observed toward the sub-mm continuum peak
of SFO 11NE SMM1. Model spectra from the best fitting line model
described in Sect.~\ref{sect:model} are
shown by dotted thin and thick lines.}
\label{fig:centrespec}
\end{figure}

The data were reduced using the Starlink package SPECX (Prestage et al.~\cite{specx}).
To improve the signal-to-noise ratio the spectra were binned to a velocity resolution
of 0.2 km\,s$^{-1}$.  The HCO$^{+}$ spectra are shown in Fig.~\ref{fig:gridspec},
overlaid on the SCUBA 850$\mu$m image from Thompson et al.~(\cite{thompson03a}). The
central HCO$^{+}$ spectrum clearly displays a double-peaked profile in both the binned
and unbinned spectra, although skewed to the red rather than the typical blue-excess
infall profile (Myers et al.~\cite{myers00}). A composite spectrum of the HCO$^{+}$
and H$^{13}$CO$^{+}$ spectra taken at the central position may be found in
Fig.~\ref{fig:centrespec}. The H$^{13}$CO$^{+}$ line has a moderate asymmetry, with a
clearly visible red wing.

\section{Modelling the dust continuum emission and gas kinematics}
\label{sect:model}

The overall appearance of the HCO$^{+}$ emission is puzzling. At the central
position of the core the HCO$^{+}$ emission is double-peaked, which is
suggestive of infall, but the emission from the red peak is greater than that
of the blue, perhaps implying that the core is expanding rather than
collapsing. The optically thin H$^{13}$CO$^{+}$ emission is singly peaked with
a central velocity at that of the self-absorption dip, ruling out the
possibility that the shape of the HCO$^{+}$ emission is caused by two unrelated
line-of-sight molecular clumps. Away from the central position the HCO$^{+}$
spectra are more symmetric, without clearly visible absorption dips. The outer
spectra are also approximately the same brightness, although the southern and
western spectra are slightly weaker than the remaining two spectra, which is
perhaps due to the fact that the molecular cloud is sharply truncated in these
directions (Thompson et al.~\cite{thompson03a}). The east and west spectra
appear to be slightly velocity-shifted from one another, which may suggest that
the core rotates about its north-south axis, although in these cases the
emission toward the centre of the core is usually symmetric rather than skewed
to the red (Zhou \cite{zhou95}).

As mentioned in the Introduction, the interpretation of infall line profiles can be
fraught with difficulty and in order to understand the kinematics of the molecular gas
comprising SFO 11NE SMM1 we carried out detailed radiative transfer modelling of  the
sub-mm continuum and HCO$^{+}$/H$^{13}$CO$^{+}$ emission. As SFO 11NE SMM1 is only
marginally elongated along the NW-SE axis in the SCUBA image and the  line emission
suggests that the core is reasonably spherical with no strong velocity gradient  we
decided to model the emission using the RATRAN one-dimensional radiative transfer code
(Hogerheijde \& van der Tak \cite{hvdt00}).

RATRAN solves for the line and dust continuum emission simultaneously. Firstly, we used
the azimuthally averaged SCUBA 850 $\mu$m profile and the spectral energy distribution
(at 60, 450 and 850 $\mu$m) published by Thompson et al.~(\cite{thompson03a}) to fix the
size, temperature and density profile of SFO 11NE SMM1.  In the modelling described in this paper we assumed  the
dust opacity model of Ossenkopf \& Henning (\cite{oh94}) with thick ice mantles and
10$^{5}$ years of grain growth. Our physical model for SFO 11NE SMM1 is that of a
spherical core of radius $8 \times 10^{17}$ cm  (0.26 pc) comprised of 10
logarithmically spaced radially symmetric shells with an inner cutoff at $5 \times
10^{16}$ cm. Given that we possess information on the spatial distribution of the dust
emission at only one wavelength (850 $\mu$m) we did not feel it was appropriate to model
the density power-law index as an additional parameter and so we assumed a density
profile  $r^{-3/2}$. 

\begin{figure}[b]
\centering
\includegraphics[angle=-90,scale=0.6]{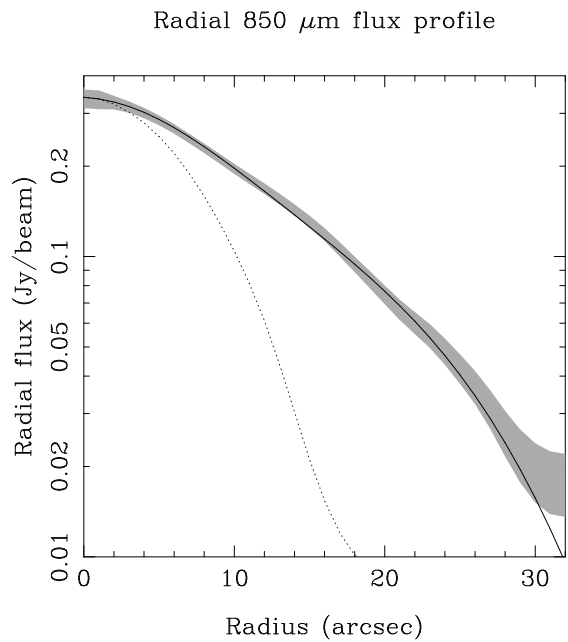}
\caption{The azimuthally averaged 850 $\mu$m flux profile and the best model fit
to the profile. The 850 $\mu$m flux profile and its uncertainty is represented by 
the shaded area. The best model fit is shown by a solid line and the JCMT beam
profile is represented by the dotted line.}
\label{fig:dustmodel}
\end{figure}

We systematically explored a wide range of parameter space to find the best model fit
to the data, including models with inwardly-increasing, outwardly-increasing and flat
(i.e.~constant) temperature gradients. In all models we assumed that the
gas and dust temperatures were equal, as is expected for high-density star-forming
cores. The model flux profile was convolved by a 14\arcsec\ FWHM beam to match the
SCUBA  observations.     The best fitting model and the observed azimuthally averaged
850 $\mu$m flux profile are shown in Fig.~\ref{fig:dustmodel}.
Fig.~\ref{fig:dustmodel} also shows that there is an upturn in the 850 $\mu$m flux at
large radii  ($>$30\arcsec), which is more than likely due to the influence of the
neighbouring core SFO 11NE SMM2 lying $\sim$50\arcsec\ to the southeast (Thompson et
al.~\cite{thompson03a}).

We could not reproduce either the observed spectral energy distribution (SED) or the
azimuthally averaged flux profile with a single-temperature model nor with an
outwardly-increasing temperature gradient. An outwardly-increasing temperature
gradient results in a much flatter flux profile than is observed, as the outer layers
start to contribute more towards the total sub-mm flux. If the dust at the outer edge
of the core is heated by the UV radiation from the nearby O-star it must comprise a
relatively thin and low-mass layer that does not substantially contribute to the 850
$\mu$m emission. By experimentation we found that the best fits to the SED and flux
profile were obtained by using a two temperature model with the inner three shells at
30 K and the remaining outer shells at 18 K. 

The best-fitting density profile for the two temperature model was found to  follow
the form  $n_{\rm H_{2}}\,(r) = 4.5 \times 10^{3}\,(r_{\rm max}/r)^{3/2}$. Both the
temperature and density compare well with the values derived from  the
single-component greybody model of Thompson et al.~(\cite{thompson03a}). The observed
flux profile and SED are reproduced well by our two-temperature $r^{-3/2}$ density
profile model. The observed 60 and 450 $\mu$m integrated fluxes are reproduced to
within 20\% and the 850 $\mu$m integrated flux to within 10\%. We estimated the
sensitivity of the best-fitting model parameters to the data by  progressively
modifying the model parameters away from the best-fitting values.  We find that the
temperature (for each component) must lie within 2 K and that the outer density 
 must lie within $3 \times 10^{2}$ cm$^{-3}$ of the best-fitting value.
 
Using the dust continuum emission to fix the size of the core and its temperature
and density profiles allows us to constrain the model of the line emission very
closely. The only free parameters in our subsequent modelling of the line emission are
the abundance of HCO$^{+}$, the mean velocity of each shell and the velocity
dispersion of the molecular gas. It is also possible to treat the $^{12}$C/$^{13}$C
abundance ratio as another free parameter but in the modelling we carried out for this
paper we assumed a standard $^{12}$C/$^{13}$C abundance ratio of 60 (Frerking, Langer
\& Wilson \cite{flw82}). 

We modelled the HCO$^{+}$ and H$^{13}$CO$^{+}$  emission from SFO 11NE SMM1 along three
broad hypotheses: pure infall, pure expansion, or a mixture of infall and expansion.
The former set of models were included as a cross-check of our method as models
comprised of pure infall should be unable to reproduce the observed red-skewed profile.
The red-skewed HCO$^{+}$ line profile suggests the presence of expanding gas which may
originate from either bulk expansion of the gas comprising the core or molecular
outflows driven by protostars embedded within the core. There is some evidence that
molecular outflows are associated with SFO 11NE SMM1, by the presence of moderate
velocity CO line wings and a possible photoionised jet  (Thompson et
al.~\cite{thompson03a}), but no other supporting evidence (e.g.~spatially separated
outflow lobes) has yet been found. Without spatially-resolved outflow lobes it is not
possible to model the gas expansion in terms of a molecular outflow and so in all the
expansion models described in this paper the expansion is modelled as a bulk expansion
of the core. Nevertheless it is worth noting that any gas expansion inferred by the
modelling may result from molecular outflows rather than a bulk expansion of the core.
We will return to this point in Sect.~\ref{sect:discuss}

As expected, it was found that models comprised of pure infalling gas could not match
the observed red-excess emission along the central line-of-sight nor the apparent
symmetry of the outer lines-of-sight. Models consisting of a purely expanding core
could not reproduce the level of asymmetry between the red and blue peaks of the
central line-of-sight spectrum. In this latter case the blue peak is always much less
bright than the red peak. For both sets of models we investigated whether velocity
gradients across the model core (e.g.~increasing infall velocity of each shell with
decreasing shell radius or vice versa) could match the observed line profiles. In all
cases the observed HCO$^{+}$ emission could not be reproduced by either pure infall or
pure expansion.

We decided to restrict the models including a mixture of infalling and expanding gas 
to a simple two-layer interpretation to simplify the modelling process and to reduce
the number of free parameters in the model. This two-layer interpretation is similar
to the simplified model described by Myers et al.~(\cite{myers00}) but with spherical
geometry rather than two uniform collapsing layers. For simplicity we will refer to
the central portions of our spherical model as the ``core'' and the surrounding outer
layers as the ``envelope''. We considered cases for both an expanding core combined
with a  collapsing envelope (e.g.~NGC 2264 IRS1; Williams \& Garland \cite{wg02});
and a collapsing core combined with an expanding envelope. In addition we modelled 
the subsets of these two cases with either static cores or envelopes.

The best fit to the observed HCO$^{+}$ and H$^{13}$CO$^{+}$ spectra (i.e.~with the
lowest residuals) was produced by a model comprised of a collapsing core surrounded by
an expanding envelope. Models consisting of an expanding core surrounded by a 
collapsing envelope produce a central blue-skewed profile and are not consistent with
the observed spectra. The limited number of free parameters in the model (HCO$^{+}$
abundance profile, mean velocity and velocity dispersion) and the sharp discontinuity
between the two layers meant that it was difficult to closely reproduce the depth of
the HCO$^{+}$ absorption dip between the red and blue peaks of the central spectrum.
Nevertheless, the collapsing core and expanding envelope model closely matches the
relative brightness of both peaks, their displacement from one another in velocity, the
moderate red asymmetry of the H$^{13}$CO$^{+}$ line and the relative symmetry of the
outer spectra.

We could not simultaneously fit any of the outlying HCO$^{+}$ spectra in the grid-map
with a single model and this probably reflects the limitations of assuming spherical
symmetry in the model. However, the collapsing core and expanding envelope model
produces the closest fit to the outer spectra of all modelled hypotheses. The models
comprised of pure infall or pure expansion show pronounced asymmetry in their outer
HCO$^{+}$ spectra. Only by combining infalling and expanding gas motions in the model
can we reduce the asymmetry in the outer HCO$^{+}$ model spectra to approach that seen
in the observed outer HCO$^{+}$ spectra. Also, given the close match of the model to the 
HCO$^{+}$ and H$^{13}$CO$^{+}$ spectra along the central line-of-sight we are
encouraged that our model interpretation is a good representation of the gas motions
within the core.

 The largest discrepancy is in the spectrum to the north of SFO 11NE SMM1, which is much broader
and brighter than the model suggests. The eastern and western HCO$^{+}$ model spectra are a much
better match to the observations, although the western spectrum has a small velocity shift toward
the  blue. The difference between the eastern and western spectra may be due to the rotation of SFO
11NE SMM1 about its north-south axis, although the difference in the centroids of the two lines  is
too small ($\sim 0.2$ km\,s$^{-1}$) to show conclusive evidence for rotation. The model offers the
best fit to the southern spectrum, reproducing the relatively broad red peak, the small possible
self-absorption dip and  the smaller blue peak visible in the observed spectrum.

\begin{figure}[t]
\includegraphics[scale=0.8, angle=-90]{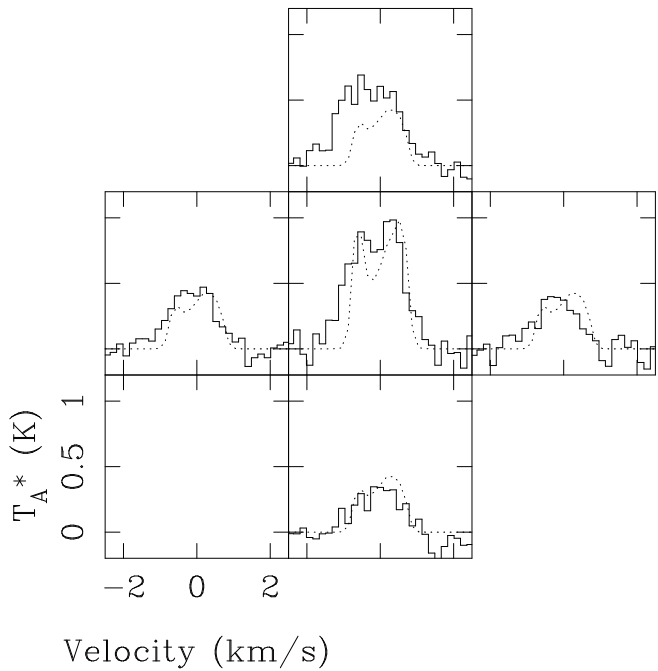}
\caption{The HCO$^{+}$ J=3--2 grid-map of SFO 11NE (\emph{solid lines}) and the
best-fitting line model (\emph{dotted lines}) described in Sect.~\ref{sect:model} .}
\label{fig:linemodel}
\end{figure}

The best-fitting model spectra of HCO$^{+}$ and H$^{13}$CO$^{+}$  towards the central
line-of-sight of SFO 11NE SMM1 are shown in Fig.~\ref{fig:centrespec}, whilst the
model HCO$^{+}$ spectra over the entire observed 5-point grid are shown in
Fig.~\ref{fig:linemodel}. The velocity field that best reproduces the observed HCO
$^{+}$ line profiles is of a central core collapsing at a constant velocity of 0.1
km\,s$^{-1}$ surrounded by an outer envelope expanding at a constant velocity of 0.3
km\,s$^{-1}$. The central core has a radius of 0.1 pc, consisting of the inner
eight model shells, and the outer envelope comprises the outer two shells of the
model.

It was found that the central HCO$^{+}$ line profile  is particularly  sensitive to
the difference in velocity between the collapsing core and expanding envelope.
Changing their relative velocities by only 0.1 km\,s$^{-1}$ is sufficient to
drastically alter the relative brightness of the red and blue peaks of the central
spectrum. The central HCO$^{+}$ line profile is also  sensitive to the relative
diameters of core and envelope. Good fits were obtained for an expanding envelope
consisting of the outer three shells of the model, changing the envelope to comprise
of the outer two layers or the outer four layers results in red peaks of the
HCO$^{+}$ central spectrum that are inconsistent with the observations.  

The HCO$^{+}$ line profiles could not be reproduced by assuming a constant HCO$^{+}$
abundance across the model and we found that smoothly increasing the  abundance of
HCO$^{+}$  from 10$^{-9}$ at the model edge to $5 \times 10^{-8}$  at the model
centre gave the best results. The velocity dispersion in the model primarily affects
the width of the resulting line profiles. We assumed a constant value of 0.3
km\,s$^{-1}$ for the velocity dispersion, which closely matches the observed
linewidths of the HCO$^{+}$ and H$^{13}$CO$^{+}$ lines. The model spectrum at the
central position is narrower than the observed spectrum and this could indicate that
the velocity dispersion increases towards the centre of the cloud. However, in the
interests of restraining the number of free parameters on the model we chose not to
explore this possibility.

\section{A collapsing molecular core with an expanding outer envelope?}
\label{sect:discuss}

In the previous section we described our physical model for SFO 11NE SMM1 as a dense
central core of radius 0.1 pc collapsing at $\sim 0.1$ km\,s$^{-1}$, surrounded by an
outer envelope expanding at a velocity of 0.3 km\,s$^{-1}$. Here, we will explore
the consequences of and offer a plausible explanation for the collapse and
expansion of the molecular cloud core.

Integrating over the entire model results in a total mass of 25 M$_{\odot}$ for SFO
11NE SMM1, which is within a factor of 2 of the greybody estimate of Thompson et
al.~(\cite{thompson03a}). As the density distribution in the model goes as
$r^{-3/2}$ and the model shells are logarithmically spaced, most of the mass lies in
the outer shells of the model. For our collapsing core and expanding envelope this
translates into the core possessing roughly a third of the total mass (8 M$_{\odot}$),
whilst the envelope consists of the remaining two-thirds (17 M$_{\odot}$).

The overall picture of SFO 11NE SMM1 gained from the modelling is of a low-mass
central collapsing core surrounded by a relatively high-mass expanding envelope. It
is unlikely that the double-peaked line profile is caused by two unrelated gas
components at different velocities (e.g.~the core and north-south ridge identified by
Thompson et al.~\cite{thompson03a}) as the optically thin H$^{13}$CO$^{+}$ line shows
a single-peak at the systemic velocity. A possible explanation for the observed line
profile is that the collapse motions result from infall onto one or more prototstars
at the centre or the core, whilst the expanding outer layers represent a
photoevaporated flow caused by the illumination of the outer layers of the cloud by
the nearby O-star.

We may use the masses of each component, their radii and respective collapse and
expansion velocities to estimate the mass infall and expansion rates to first order,
allowing us to investigate this hypothesis. The simple relation $\dot{M} = M\,v/r$,
relates the mass infall or expansion rate $\dot{M}$ to the mass of the core
or envelope ($M$),  the infall or expansion velocity ($v$) and the outer radius of
the  core or inner radius of the envelope ($r$). The infall rate $\dot{M}_{\rm in}$
estimated from this relation is $\sim 7 \times 10^{-6}$ M$_{\odot}$\,yr$^{-1}$ and
the corresponding  expansion rate $\dot{M}_{\rm exp}$ is $\sim 4 \times 10^{-5}$
M$_{\odot}$\,yr$^{-1}$.

The infall rate inferred from the model is consistent with a purely gravitational
flow or free-fall collapse, i.e.~ $\dot{M}_{\rm in} \simeq \sigma^{3}/G$, where
$\sigma$ is the velocity dispersion (predicted to be 0.3 km\,s$^{-1}$ from the model)
and $G$ is the gravitational constant (Williams \& Garland \cite{wg02}). The infall
rate is also approximately equal to that that estimated for a roughly solar-mass
protostar (Reid, Pudritz \& Wadsley \cite{rpw02}; Zhou et al.~\cite{zhou95}). Is it
possible that we are observing infall onto a single protostellar object at the 
centre of SFO 11NE SMM1?

2MASS near-infrared images reveal a candidate cluster of class I protostars and YSOs
at the heart of SFO 11NE SMM1 (Thompson et al.~\cite{thompson03a}) and so it is
possible that we are observing gas that is either collapsing onto the cluster as a
whole or perhaps onto individual protostars within the cluster. Hydrodynamical models
of cluster formation suggest that as the cluster forms and contracts the steepening
gravitational potential funnels an accreting gas flow onto the cluster (Bonnell et
al.~\cite{bcbp01}). We note, however, that the mass infall rate within SFO 11NE SMM1
is much lower than has been observed for other protostellar clusters (e.g.~NGC 2264
IRS1; Williams \& Garland \cite{wg02}), where the infall rate can be up to  2 orders
of magnitude greater than that for single protostars. For the gas to be  undergoing a
large scale collapse onto the cluster as a whole there must be a retarding mechanism
in operation to slow the infall rate to that consistent with a single protostar.

One of the predictions of radiative-driven implosion (RDI) models of bright-rimmed
clouds is the presence of a photoevaporated flow from the surface of the cloud
illuminated by the nearby OB star (Bertoldi \cite{bertoldi}; Lefloch \& Lazareff
\cite{ll94}). The expanding outer layers of the cloud, may in part be due to the
presence of a photoevaporated flow from the surface of the cloud, although the large
mass (2/3 of the total molecular cloud core mass) and size of the expanding layer are
somewhat difficult to reconcile with a photoevaporated flow alone.  The mass loss
rate from SFO 11NE SMM1 caused by the photoevaporated flow from the cloud surface 
was estimated by Thompson et al.~(\cite{thompson03a}) to be  approximately 3.5
M$_{\odot}$\,Myr$^{-1}$, which is roughly an order of magnitude lower than that
implied by the mass and expansion velocity of the outer envelope ($\sim 40$
M$_{\odot}$\,Myr$^{-1}$).

The relatively high mass expansion rate implies
that the envelope expansion is unlikely to be solely driven by a photoevaporative flow.
There are two alternative hypotheses that may explain the presence of
expanding gas in within SFO 11NE SMM1: either  molecular outflow(s) from protostars
embedded within SFO 11NE SMM1 or a bulk re-expansion of the bright-rimmed cloud. 
The former possibility of outflow was mentioned in the previous section: the expanding
envelope of the model may be due to one or more molecular outflows mimicking the
signature of the bulk expansion of the core. 

However, we do not see any evidence for the high-velocity line wings that are the usual
indicators of molecular outflow in either the HCO$^{+}$ spectra contained in this paper
nor in the CO spectra of Thompson et al.~(\cite{thompson03a}). Although the model of the
cloud does not specifially consider a bipolar flow, we would expect the predicted mass
of the expanding gas to agree with that contained in a bipolar outflow to at least first
order. The mass of outflowing gas predicted by the model (17 M$_{\odot}$) is much
greater than that typically observed in low mass outflows, which is usually a few
M$_{\odot}$ (e.g.~Fukui et al.~\cite{fukui93}; Richer et al.~\cite{richer00}). It is
possible that the mass inferred by the model could be comprised of multiple overlapping
flows, particularly in regard of the candidate cluster of protostars and YSOs found at the
heart of SFO 11NE SMM1. Although outflows cannot be completely ruled out as an
underlying cause of the gas expansion, the lack of supporting evidence  for the presence
of outflows in this bright-rimmed cloud (for example spatially resolved outflow lobes)
and the high outflow  masses required to reproduce the model predictions substantially
undermine the likelihood of this hypothesis.

Another possibility is that the expanding gas is caused by a bulk re-expansion of the
cloud. RDI models predict that this effect follows the maximum compression of the cloud
by photoionisation-induced shocks. As the maximum compression is reached, the cloud
overshoots its equilibrium state and subsequently expands (Lefloch \& Lazareff
\cite{ll94}). Thereafter the cloud follows an oscillating and decaying cycle of
compression and re-expansion until the gas reaches a quasi-static hydrodynamic
equilibrium phase as a cometary cloud. Perhaps the expanding outer layers of
the model cloud correspond to this expansion phase.

Lefloch \& Lazareff define a ``radial momentum'' for their RDI model clouds which is
the integral of the gas momentum over the cloud volume. The radial momentum of our
expanding envelope (0.8 M$_{\odot}$\,pc\,Myr$^{-1}$) is well within the angular
momentum limits evaluated by Lefloch \& Lazareff (\cite{ll94}), even taking into
account the relatively high mass fraction of the envelope to central core. Thus it
seems that the expanding outer envelope is consistent with the re-expansion phase of
the RDI models of bright-rimmed clouds, which is suggested to occure approximately .  
$3 \times 10^{5}$ years after the initial exposure of the cloud to ionising radiation
(Lefloch \& Lazareff \cite{ll94}). This value is consistent with the upper bound
on the cloud age derived from the projected distance to the ionisation bound of the
HII region (Thompson et al.~\cite{thompson03a}).  


On balance, given the lack of supporting evidence for molecular outflows in SFO 11NE
SMM1 we conclude that the most likely cause of the gas expansion is a genuine bulk
expansion of the cloud following the maximum compression of the cloud by
photoionisation-induced shocks. The likelihood is that we are seeing SFO 11NE SMM1 in
the beginning of its re-expansion stage, as this stage is the most consistent with the
dynamic age of the associated HII region and the duration that the cloud has been
exposed to the UV radiation from the nearby OB star. However, in order to confirm this
hypothesis further observations aimed at confirming or denying the presence of bipolar
molecular outflow are a high priority.


\section{Summary and conclusions}

We observed the bright-rimmed cloud core SFO 11NE SMM1 in sub-millimetre wave lines
of HCO$^{+}$ and H$^{13}$CO$^{+}$ to search for the characteristic optically thick
blue-peaked infall signature (e.g. Myers et al.~\cite{myers00}). The HCO$^{+}$
emission toward SFO 11NE SMM1 is indeed double-peaked, but with an excess toward the
red end of the spectrum rather than the blue. The H$^{13}$CO$^{+}$ spectra show a
single peak at the systemic velocity and this suggests that the double-peaked
optically thick profile originates from a single gas core rather than a line-of-sight
coincidence of two different velocity components.

The line data were modelled with the  radially symmetric radiative transfer model
RATRAN (Hogerheijde \& van der Tak \cite{hvdt00}). SCUBA and IRAS HIRES data from
Thompson et al.~(\cite{thompson03a}) were used to construct a detailed physical model
of the temperature and density structure of the cloud core against which the line
emission could be modelled. The resulting physical model of the source is a cloud core of
radius 0.26 pc, with a density at the outer radius of $4.5 \times 10^{3}$ cm$^{-3}$
increasing toward the centre of the core as r$^{-3/2}$. The SED and radial flux profile of the
core suggest that the temperature of the core consists of two components: an inner
warm region at 30 K surrounded by a colder envelope of 18 K.

The HCO$^{+}$ and H$^{13}$CO$^{+}$ line profiles were found to be consistent with a
combination of a central core collapsing at a constant velocity of 0.1 km\,s$^{-1}$, 
surrounded by an expanding envelope moving outwards at a constant velocity of 0.3
km\,s$^{-1}$.  We derive the mass infall and expansion rates to first order  by a simple
consideration of the mass, velocity and radius of the collapsing and expanding
regions. The mass infall rate is consistent with that estimated for a single
solar-mass protostar (e.g.~Zhou \cite{zhou95}) and it is possible that we are
observing collapse onto a single member of the infrared cluster seen at the heart of
SFO 11NE SMM1 (Thompson et al.~\cite{thompson03a}).  

The expansion rate is an order of magnitude larger than can be sustained by a
photoevaporative flow from the outer layers of the cloud (Lefloch \& Lazareff
\cite{ll94}). The expansion is consistent with either the bulk re-expansion of the cloud
predicted by the RDI models of Lefloch \& Lazareff (\cite{ll94}) or multiple unresolved 
bipolar molecular outflows. Given both the lack of supportive evidence for outflows and
the relatively high mass of expanding gas predicted by the radiative transfer modelling
we conclude that the likeliest scenario for the expanding gas is a bulk re-expansion of
the bright-rimmed cloud. If so, SFO 11NE SMM1 is at the beginning of its re-expansion
phase following the maximum compression of the cloud by photoionisation-induced shocks.

Although our model of the cloud core as undergoing central collapse and outer
expansion is consistent with the data, we stress that this is just one possible
interpretation. Observations of higher sensitivity and  spatial resolution are
required to investigate the kinematics of SFO 11NE SMM1 at smaller spatial scales and
provide sufficient data for fully two-dimensional radiative transfer  modelling of the
line profiles (e.g.~Hogerheijde \& van der Tak \cite{hvdt00}; Phillips \& Little
\cite{pl00}) . In particular, the possibilities that the cloud core is rotating about
its north-south axis and that the observed expansion may be in part due to bipolar
molecular outflows must be investigated. Our forthcoming survey of a statistically
significant number of BRCs will also reveal whether infall motions are as common
within these clouds as more isolated low-mass star-forming regions.


\begin{acknowledgements}   
The authors would like to thank Larry Morgan for struggling
through his altitude-induced adversity to assist with the observations, Antonio
Chrysostomou for useful discussions,  and the JCMT staff for a pleasant and productive
observing run. MAT is supported by a PPARC postdoctoral grant.  
\end{acknowledgements}

\end{document}